\documentstyle[aps,twocolumn,epsf,tighten]{revtex}
\begin{document}
\title
{\hfill
\begin{minipage}{0pt}\scriptsize \begin{tabbing}
\hspace*{\fill} GUTPA/00/04/02\\
\hspace*{\fill} WUP-TH 00-13\\ 
\hspace*{\fill} HLRZ2000-11 \end{tabbing} 
\end{minipage}\\[8pt] 
A high precision study of the $Q\bar{Q}$ potential from Wilson loops\\
in the regime of string breaking
}
\author{Bram Bolder$^{a}$, Thorsten Struckmann$^{b}$\\
Gunnar S.\ Bali$^{c}$, Norbert Eicker$^{a}$,
Thomas Lippert$^{a}$, Boris Orth$^{a}$,
Klaus Schilling$^{a,b}$, and  Peer Ueberholz$^{a}$}
\address{${}^{a}$Fachbereich Physik, Bergische Universit\"at, Gesamthochschule
Wuppertal\\Gau\ss{}stra\ss{}e 20, 42097 Wuppertal, Germany}
\address{${}^{b}$NIC, Forschungszentrum J\"ulich, 52425 J\"ulich and\\
DESY, 22603 Hamburg, Germany}
\address{${}^{c}$Department of Physics and Astronomy, 
The University of Glasgow,
Glasgow G12 8QQ, Scotland\\[.2 cm]
(SESAM - T$\chi$L  Collaboration)}
\maketitle
{
\begin{abstract}
  For lattice QCD with two sea quark flavours we compute the static quark
  antiquark potential $V(R)$ in the regime where string breaking is expected.
  In order to increase statistics, we make full use of the lattice information
  by including {\it all} lattice vectors ${\mathbf R}$
  to any given separation $R=|{\mathbf R}|$
  in the infrared regime. The corresponding paths between the lattice points
  are constructed by means of a generalized Bresenham-algorithm as known
  from computer 
  graphics. As a result, we achieve a determination of the unquenched
  potential in the
  range $0.8$ to $1.5$ fm with hitherto unknown precision.
  Furthermore, we demonstrate some error reducing
  methods for the evaluation of the transition matrix element between two- and
  four-quark states.


\end{abstract}

\narrowtext
\section{Introduction}
Confinement of quarks is an issue of prime importance in the understanding of
strong interaction physics. While the study of the static 
quark-antiquark potential from simulations of quantum chromodynamics has been
pushed to rather high accuracy in quenched QCD and allows today for a precise
determination of the string tension, the search for evidence of string
breaking from Wilson loops in simulations of the full QCD vacuum has been
futile so far. It seems that the linear rise of  the static potential
continues to 
prevail even in presence of vacuum polarization by quark loops~
\cite{Heller:1994rz,Glassner:1996xi,Bali:1998bj,Aoki:1999sb,Garden:1999hs,Bernard:2000gd}.

The common explanation for this unexpected finding is that present studies
cannot really resolve the asymptotic time behaviour of the Wilson loops and
that a multichannel analysis including light fermion operators is required to
achieve sufficient ground state overlap in the available time
range~\cite{Gusken:1998sa}.  In fact, a fully fledged multichannel approach
has been demonstrated to be a viable technique to realize breaking of the
string between adjoint sources in pure gauge theories or fundamental colour
sources in Higgs
models~\cite{Knechtli:1998gf,Philipsen:1998de,Stephenson:1999kh}. In the case
of full QCD, however, it appears overly costly to achieve the required
statistical accuracy of the generalized Wilson loops which incorporate light
quark-antiquark pairs in the initial or final states~\cite{Schilling:1999mv}.
The reason is that one is prevented from exploiting the translational
symmetry, since this would require computation of light propagators $P(y,x)$
on any source point location, $x$ (see e.g.\ 
Refs.~\cite{Schilling:1999mv,Bali:2000gf}). In
Refs.~\cite{Lacock:pisa,Pennanen:2000yk} stochastic estimator methods with
maximal variance reduction (so-called {\it all-to-all methods}) were applied
to cope with the fluctuations on the multichannel correlator matrix, but
failed so far to provide sufficient accuracy in the infrared regime.

On the other hand, if one scrutinizes existing QCD potential data from Wilson
loops~
\cite{Heller:1994rz,Glassner:1996xi,Bali:1998bj,Aoki:1999sb,Garden:1999hs,Bernard:2000gd}
for colour screening one will notice that the errors become substantial in the
region of interest, $r \simeq 1.2 $ fm\footnote{We use capital (lower case)
  letters for quantities in lattice units (physical units).}. Thus, there is
room for suspicion that QCD colour screening has so far escaped detection
simply for the lack of adequate precision of large Wilson loops.

The main purpose of the present note is to improve on this point by pushing
for a high precision `classical' Wilson loop calculation at large separations
in $N_f = 2$ QCD. This is achieved by squeezing maximal information out
of each vacuum configuration
through rotational invariance and comprehensive utilization  of {\it all
possible $R$-values} on the lattice.  As a result of our ``all-$R$ approach''
(ARA) we are able to present a long range static potential from Wilson loops
in unprecedented accuracy.

In section \ref{section:loops} we shall describe how we go about in the
build-up of nonplanar loops to any given ${\mathbf R}$. Section
\ref{section:potential} contains our potential analysis from ARA which is used
on top of existing signal enhancement  techniques, such as conventional APE
smearing\cite{Albanese:1987al} and translational averaging.  As a first step
in the direction of a two channel investigation of string breaking we study in
section~\ref{section:noise} the noise reduction effect from ARA on the
transition correlator between static and static-light quark states, $\bar{Q}Q$
and $\bar{Q}Q\bar{q}q$, respectively.

\section{Loop construction}
\label{section:loops}
Our aim is to increase statistics in the regime of colour screening,
i.e. for large quark antiquark separations, $R$.  Obviously, on such
a length scale, a given QCD vacuum configuration contains plenty of
information that can be exploited for self averaging and thus for
error reduction: firstly, one can realize, on a hypercubic lattice, a
fairly dense set of $R$ values; secondly, for a given   large value
of $R = |{\mathbf  R}|$, there are many different three-vector realizations
${\mathbf R}$ on the lattice.

We wish to make use of this fact by a systematic construction of off-axis
Wilson loops, $W(R,T)$, in the range $R_{\min} \leq R \leq R_{\max}$, with
$R_{\min} = 10\approx 1.7\, R_0$ and $R_{\max} = 12\sqrt{3}\approx 3.5\, R_0$, 
where $R_0$ is the Sommer radius (in
lattice units) that amounts to $r_0\approx 0.5$ fm~\cite{Sommer:1994ce}.  The
construction proceeds by choosing all possible vectors, ${\mathbf R}$, with
integer components $C_{\min}$, $C_{\mbox{\scriptsize mid}}$, and $C_{\max}$
(in any order of appearance)
that obey the inequalities,
\begin{equation}
R_{\min}^2 \leq R^2 =
C_{\min}^2 + C_{\mbox{\scriptsize mid}}^2 + C_{\max}^2  \leq R_{\max}^2 \; ,
\label{eq:inequality}
\end{equation}
where $|C_{\min}| \leq |C_{\mbox{\scriptsize mid}}|\leq |C_{\max}|$.

Subsequently, the set of solutions to the constraint eq.~(\ref{eq:inequality})
is sorted according to the correponding values of $R$.  In
Table~\ref{table:combis} we display the large number of possible $R$ values
and vectors, ${\mathbf R}$, obtained in this way, with the additional
restriction, $|C_{\max}|\leq 12$.  So far, only ${\mathbf R}$ vectors with
$(|C_{\max}|,|C_{\mbox{\scriptsize mid}}|,|C_{\min}|)$ being multiples of
$(1,0,0), (1,1,0), (1,1,1), (2,1,0), (2,1,1)$ or $(2,2,1)$ have been
considered~\cite{Bali:1992ab,Bali:1993ru,Bali:1997am,Bali:2000vr}.  Within the
investigated regime, $10 \leq R \leq 12 \sqrt{3}$, we achieve a gain factor of
more than eight in terms of the spatial resolution in $R$ (from 21 to 175
different values).  Moreover, the number of different ${\mathbf R}$ vectors
that yield the same distance, $R$, is increased by an average gain factor of
more than four!   Accordingly, we   find the ARA to reduce the
statistical errors on potential values by factors around two.

We shall briefly discuss the construction of the gauge transporters connecting
the quark and antiquark locations, ${\mathbf R}_Q$ and ${\mathbf R}_{\bar{Q}}
= {\mathbf R}_Q + {\mathbf R}$, that appear within the ARA nonplanar Wilson
loops.  In order to achieve a large 

\begin{table}
\caption{The number of
different solutions to eq.~(\protect\ref{eq:inequality})
obtained by ARA for the interval
$10\leq R\leq 12\protect\sqrt{3}$, $|R_i|\leq 12$.}
\label{table:combis}\vskip .2cm
\begin{tabular}{c|c|c|c}
&$\# R$ values&$\# {\mathbf R}$ vectors&$\# {\mathbf R}$ vectors/$R$-entry\\\hline
           standard & 21 & 302 & 14.4 \\
           ARA      & 175& 11486 & 65.6
\end{tabular}
\end{table}
\noindent
overlap with the physical ground state we
would like to construct lattice paths that follow as close as possible the
straight line connecting ${\mathbf R}_{\bar{Q}}$ with ${\mathbf R}_Q$.  This
task can be accomplished by a procedure which is known as the Bresenham
algorithm~\cite{bresenham} in computer graphics. There one wishes to map a
straight continuous line between two points, say ${\mathbf 0}$ and ${\mathbf
  C} = (C_{\max}, C_{\min})$, onto discrete pixels on a 2-d screen.  Then one
has to find the explicit sequence of pixel hoppings in max- and min-directions
such that the resulting pixel set mimicks best the continuum geodesic between
points ${\mathbf 0}$ and ${\mathbf C}$. The Bresenham prescription is simply
to move in max-direction unless a step in min-direction brings you closer to
the geodesic from ${\mathbf 0}$ to ${\mathbf C}$, where we assume
$|C_{\max}|\geq |C_{\min}|$.  It can easily be embodied into a fast algorithm
based on local decision making only (see Fig.~\ref{fig:bresenham2}), by means
of a characteristic lattice function $\chi$ that incorporates the aspect ratio
$C_{\max}/ C_{\min}$.

\begin{figure}
\centerline{\epsfxsize=4cm\epsfbox{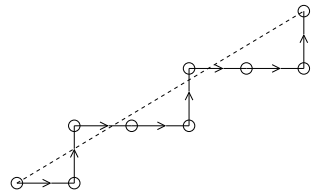}}
\vskip .4 cm
\caption{Illustration of a path construction by the Bresenham algorithm,
for ${\mathbf C} = (5,3)$.}
\label{fig:bresenham2}
\end{figure}
The algorithm in two dimensions  looks like this:\\[.4cm]

\noindent
{\sf
cmax2 := 2*cmax\\
cmin2 := 2*cmin\\
chi := cmin2 -cmax\\
FOR i := 1 TO cmax DO
\begin{quote}
\vskip -.2cm
{\it step in max-direction}\\
IF chi $\geq$ 0 THEN
\begin{quote}
\vskip -.2cm
chi  := chi - cmax2\\
{\it step in min-direction}
\end{quote}
\vskip -.2cm
ENDIF\\
chi := chi + cmin2
\end{quote}
\vskip -.2cm
ENDDO
}\\[.5 cm]
\noindent
The generalisation to three dimensions is achieved by
combining two of these 2d-algorithms with different $\chi$'s for max-mid and
max-min in just one loop over the max-direction.  

In order to convey an idea
about the statistics gain inherent in such a systematic approach we have
listed the number of $ {\mathbf R}$ vectors   constructed in this manner in
Table~\ref{table:combis}. Note that for plane or space diagonal separations,
we average over the two or six equivalent paths, respectively.

\section{The static potential at large $R$}
\label{section:potential}
We base our analysis on 184 vacuum configurations,
separated by one autocorrelation length,  on $
L_{\sigma}^3L_{\tau}=24^3 \times
40$ lattices of $N_f =2 $ QCD at $\beta = 5.6$ and $\kappa_{sea} = .1575$
(corresponding to $m_{\pi}/m_{\rho} = .704(5)$),
produced by the T$\chi$L-collaboration on an APE~100 tower at INFN.
These parameter values correspond to the biggest
lattice volume at our disposal, $L_{\sigma}a\approx 2$~fm.
In order to
minimize finite-size effects and possible violations of rotational
symmetry on the $L_{\sigma}=24$ torus,
$C_{\max}$ has been restricted to $|C_{\max}|\leq 12$.
The lattice constant $a$ was determined from the Sommer radius
$r_0=R_0a\approx 0.5$~fm~\cite{Sommer:1994ce}, as obtained in our previous
investigation~\cite{Bali:2000vr}, $R_0 = 5.89(3)$.

Before entering the Wilson loop analysis the configurations are smoothened by
spatial APE- or link smearing~\cite{Albanese:1987al}, 
\begin{equation}
\label{eq:fuzzing}
\mbox{link} \rightarrow \alpha \times \mbox{link} + \mbox{staples} \; ,
\end{equation}
followed up by a projection back into the gauge group~\cite{Bali:1992ab}, with
26 such iterative replacements and the parameter value, $\alpha = 2.3$.

The potential values have been obtained by means of single and double
exponential fits to smeared ARA Wilson loop data $W(R,T)$
within the range, $T_{\min}\leq T\leq 8$.
We shall quote statistical errors that are obtained
by jackknifing. 
At the large $R$ values
that are of interest in view of screening effects, the quality of the
statistical signal did not allow to include $T$ values larger than $8$.

\begin{figure}[b]
\centerline{\epsfxsize=9cm\epsfbox{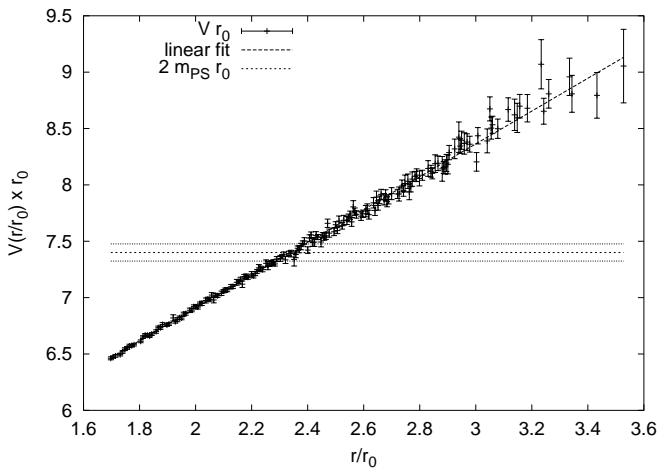}}
\vskip .4 cm
\caption{The estimate on the static potential as obtained from
Wilson loops at $\kappa = 0.1575$
and $\beta = 5.6$ on $24^3 \times 40$ lattices, extracted
from a single exponential fit to Wilson loops for $4 \leq T \leq 8$.}
\label{fig:potential}
\end{figure}
In Fig.~\ref{fig:potential} we display estimates on the potential in the range
$1.7 \leq r/r_0 \leq 3.5$ obtained from a single exponential fit with
$T_{\min}=4$ (they agree with results from a double exponential fit with
$T_{\min}=1$). Note that for $r> 2.4\,r_0$ the data are to be interpreted as
strict upper limits on the potential. The slope is in agreement with the
string tension, $K=\sigma a^2=0.0372(8)=1.139(4)R_0^{-2}$, as quoted in
Ref.~\cite{Bali:2000vr} from a fit to data obtained at smaller $r<2.04\,r_0$.
Around the separation $r_{c} \approx 2.3\,r_0$ the potential energy crosses
the expected threshold for string breaking, $2m_{PS}\, a = 1.256(13)$, which
is indicated by a horizontal error band.  The errors quoted are statistical
and are well
%
%
%
below 1.5 \% for  $ r/r_0 \leq 3$.
We find the data to follow perfectly a straight line: a flattening of the
Wilson loop potential is not visible within the accessible $T$ range  and
present statistical errors.

To complement this result, we computed the overlaps with the ground and first
excited states by means of two-exponential fits, as displayed in
Fig.~\ref{fig:overlap}. Again our two data sets exhibit linear dependencies on
$r$, with nearly opposite slopes such that their sum turns out to be close to
{\it one}. The remainder does only slightly depend on $r$ and is of order 10
\% .

We conclude that Wilson loop operators are definitely not very well suited to
uncover string breaking.  To achieve the required overlap with four-quark
states one has to introduce them explicitely into the calculation, in form of
a coupled channel analysis.  As a first step in such a program, we shall
explore in the remaining part of this letter a number of signal enhancement
techniques based on ARA, in application to the transition matrix element
between two- and four-quark states.

\begin{figure}
\centerline{\epsfxsize=9cm\epsfbox{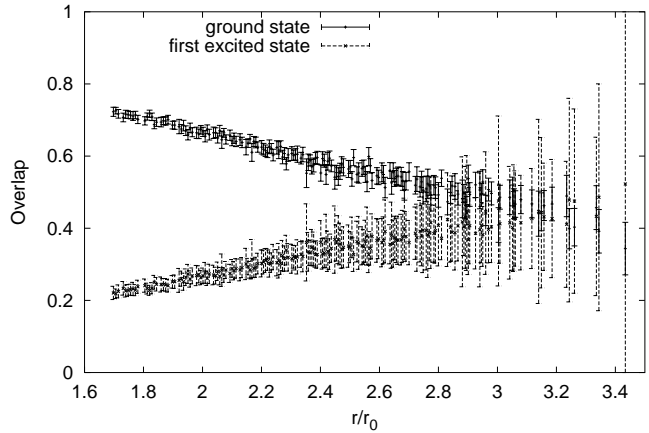}}
\vskip .4 cm
\caption{The ground and first excited state overlaps (top and lower
  data sets, respectively) plotted versus the quark-antiquark separation, as
  obtained from Wilson loops. The data result  from two-exponential fits in
  the region $1\leq T \leq 8$.}
\label{fig:overlap}
\end{figure}

\section{Noise on the transition operator}
\label{section:noise}
In a two-channel approach one  extends the Wilson loop vacuum
expectation value, $C_{11}$, to a $2 \times 2$ correlation matrix $C$
as pictogrammed in Fig.~\ref{fig:correlation_matrix}.  The task is then to
solve the generalized eigenvalue problem~\cite{Luescher:1990ml},
\begin{equation}
\label{eq:transfer}
 C(t+\tau) {\mathbf u}_i =  \lambda_i(\tau) C(t) {\mathbf u}_i \; ,
\end{equation}
where the eigenvalues connect to the two energy levels $E_i$ at large enough
$t$:
\begin{equation}
 \lambda_i(\tau) = \exp{(-E_i\tau)} \; .
\end{equation}
Unless $C$ happens to be diagonal, the physical states ${\mathbf u}_0$ and
${\mathbf u}_1$ are mixtures of two- and four-quark states.  It is obvious
that the energy levels are insensitive to changes in the normalization of the
wavefunctions, ${\mathbf u}_i$.  Thus, in addition to spatial APE link
smearing, source smearing techniques as known from spectrum calculations are
applicable (see Fig.~\ref{fig:smearing}) and will be tested.

\begin{figure}
\centerline{\epsfxsize=2cm
$C=\left(\begin{array}{c}\epsfbox{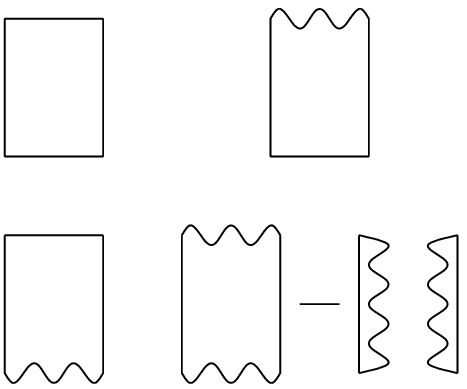}\end{array}\right)$}
\vskip .4 cm
\caption{The two channel correlation matrix $C$. Wiggly lines correspond
to light quark propagators, solid lines to products of gauge links.}
\label{fig:correlation_matrix}
\end{figure}

While the noise level on $C_{11}$ could be greatly reduced by applying ARA on
top of standard volume self averaging (VSA), the quark propagators $P(y,x)$
(wiggly lines) entering the remaining components of the correlation matrix $C$
prevent the direct use of VSA that requires matrix inversions on {\it all}
possible source points, $x$.  Pennanen and Michael have
tested noisy estimator techniques on $P$
for curing this problem but did not succeed to  reach the accuracy
necessary
to solve eq.~(\ref{eq:transfer})~\cite{Pennanen:2000yk}.
This motivates us
to explore noise controlling
strategies based on ARA rather than VSA
on the transition matrix element $C_{21}(R,T)$.
One should keep in mind that ARA can
be put to work at little extra cost since {\it one} inversion on a {\it
  single} source $x$ renders $P(y,x)$ for {\it all} sink positions $y$.
\begin{figure}[t]
\centerline{\epsfxsize=4cm\epsfbox{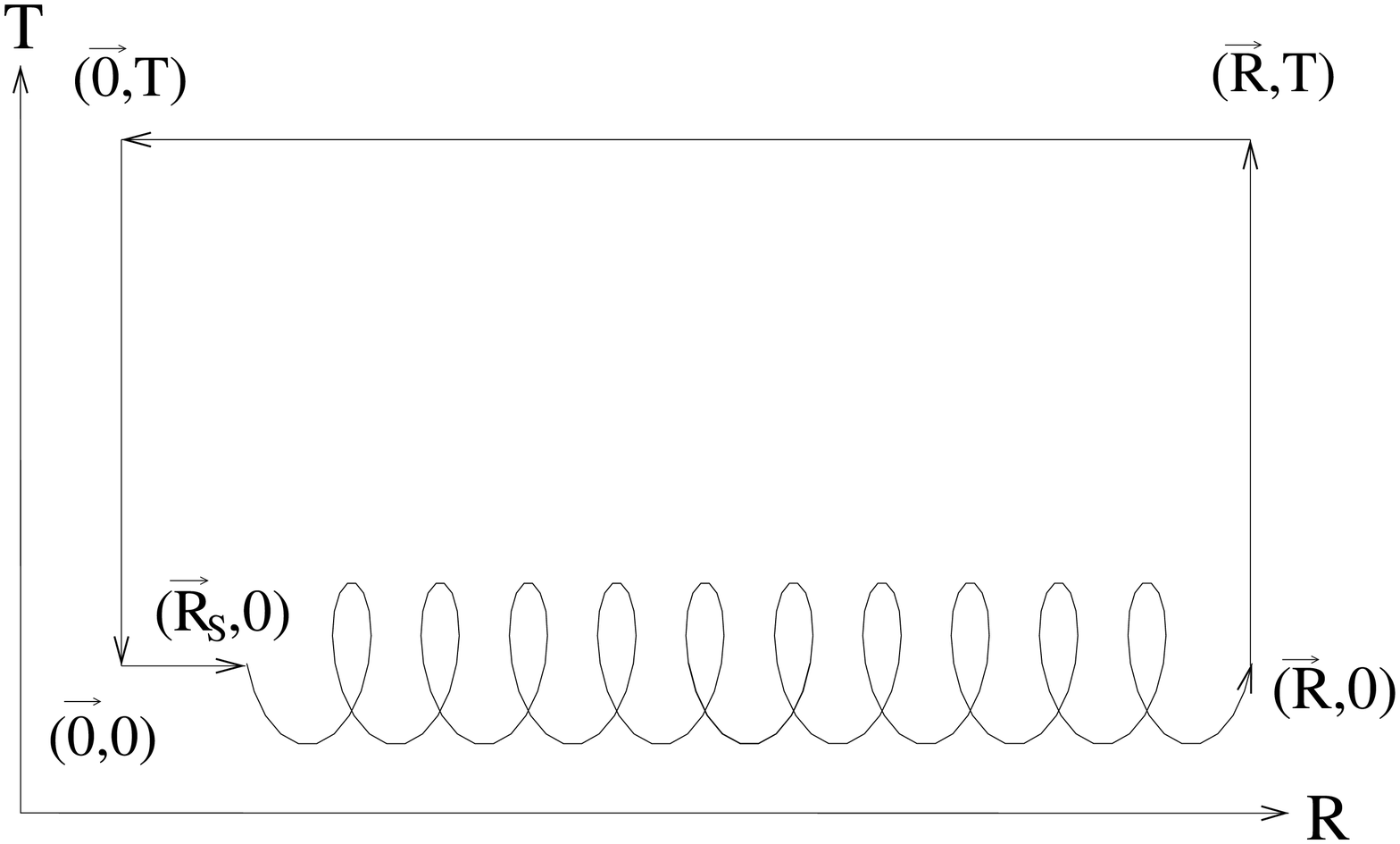}}
\vskip .4 cm
\caption{The source smeared transition matrix element $C_{21}$.}
\label{fig:smearing}
\end{figure}
In addition to APE link smearing, we have explored source smearing as
illustrated by the diagram, Fig.~\ref{fig:smearing}. This is applied  to the
light propagator source, $\chi_x$, and consists of iterative replacement (as in
previous heavy-light spectrum analyses~\cite{Guesken:1990gg,Eicker:1999sy}),
\begin{equation}
\label{eq:smearing}
\chi_x\rightarrow\chi_x+\alpha\sum_{i=\pm 1}^{\pm 3}
\chi_{x + \hat{\imath}}^{\mbox{{\tiny parall.transported}}} \; ,
\end{equation}
which is repeated 50 times,
with weight factor, $\alpha = 4.0$. The source is put to zero subsequently
outside the volume $|{\mathbf r}_s - {\mathbf x}| \leq R_q a $,
with $R_q = 5$.  Note
that quark propagators are computed without link smearing throughout this work.

Fig.~\ref{fig:noise_t1} shows the $R$-dependence of the relative errors on $
C_{21}(R, T=1)$ throughout the stringbreaking region and illustrates the
performance of various signal enhancement tools
that are added on top of ARA: (a) the
circles correspond to no source or link smearing;
(b) crosses refer to local sources and  smeared links;
(c) stars refer to smeared quark sources and smeared gauge sinks.
\begin{figure}
\centerline{\epsfxsize=9cm\epsfbox{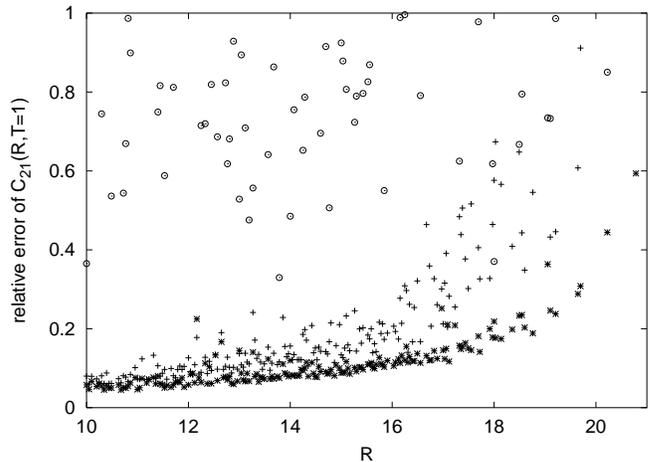}}
\vskip .4 cm
\caption{Relative errors on $C_{21}(R, T=1)$  vs.\ $R$
with ARA. String breaking is expected around $R = 13$.
Open circles: no link or source smearing; upright crosses:
link smearing only; stars: smeared source and smeared gauge links.}
\label{fig:noise_t1}
\end{figure}

Obviously link smearing helps a lot but leaves us still with errors of order 50 \%
in the region $R \leq 18$. It turns out that smearing is capable of cutting
down noise amplitudes further. We find an additional reduction of 
more than  a factor two, to a 20 \% level at $T = 1$. 

Unfortunately, such accuracy does not suffice, as it cannot be sustained at
larger $T$ values, where one wishes to analyse $C$ at the end of the day.
Given the dense set of $R$-values available from ARA and in view of the fact
that $B(R)=\ln[C_{21}(R,T)]$ is a smooth function of $R$,
there is opportunity to further improve by filtering the
sequence, $\lbrace B(R_i) \rbrace \rightarrow \lbrace B_f(R_i) \rbrace $.  We
have chosen a filter that weighs each individual jackknife sample with the
fluctuation calculated on the entire data set,
\begin{equation}
\label{eq:filter}
B_f(R_i) = N_i^{-1}  \sum_{|R_j - R_i | \leq R_f}\sigma_j^{-2}B(R_j) \; ,
\end{equation}
with normalization
\begin{equation}
\label{eq:filter_norm}
 N_i =  \sum_{|R_j - R_i | \leq R_f}\sigma_j^{-2} \; .
\end{equation}
We found best results with filter radius $R_f = 0.5$.  Let us now consider the
transition matrix element at $T=5$.  The effect
of filtering is illustrated in Fig.~\ref{fig:noise_filtering} where we
confront the signals from ARA plus link and source smearings, $\lbrace B(R_i)
\rbrace $, (upper sequence with large error bars) with the filtered one,
$\lbrace B_f(R_{{i}}) \rbrace $, (lower data sequence). In order to display
the systematics of windowing, we have refrained from thinning the latter
sequence as one should in actual applications. We find a striking noise
reduction through filtering within the large distance regime, $10 \leq R
\leq 18$: at $T = 3$ and $T =5 $ we encounter errors of less than 10 \% and 20
\%, respectively.

\begin{figure}
\centerline{\epsfxsize=9cm\epsfbox{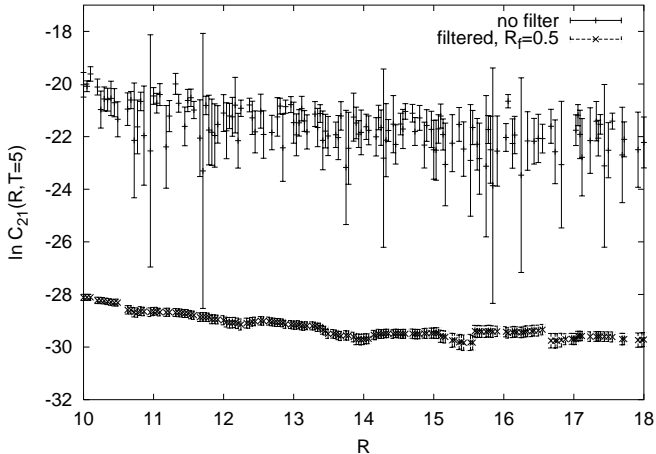}}
\vskip .4 cm
\caption{The signal for $\ln C_{21}$ at $T = 5$.
  The filtered data have been vertically shifted by $-8$ to avoid cluttering.
String breaking is expected around $R = 13$. }
\label{fig:noise_filtering}
\end{figure}

\section{Summary and Conclusions}
Exploiting the dense set of $R$ values available at large quark-antiquark
separations on the lattice we succeeded in improving the precision of the
$Q\bar{Q}$ potential from Wilson loops in the string breaking regime by a
factor of {\it two} with respect to standard methods. This enables us to
analyse Wilson loop data well beyond the point where string
breaking is expected and corroborates previous conjectures that Wilson loops
do not bear enough overlap with $Q\bar{Q}q\bar{q}$ states for uncovering
string breaking within the $T$-range available at present.
 
The success of our all-$R$ approach to the Wilson loops in the string
breaking regime has encouraged us to carry out a feasibility study on error
control of the transition correlator, $C_{21}$.  By additional use of source
smearing and filtering techniques, we arrive at reasonable signal-to-noise
ratios for $\ln C_{21}$.

For the final chord in a full two-channel analysis of $Q\bar{Q}$ and
$Q\bar{Q}q\bar{q}$ states one would have to consider the correlator $C_{22}$
which contains both, a connected and a disconnected
contribution: the former can be
tackled with our present techniques and we expect sufficient accuracy with a
factor of $T_{max}$ more effort than for $C_{21}$; the latter relates to the
situation of $B\overline{B}$ pairs at large separations.  From a previous $B
\bar{B}$ study~\cite{Pennanen:1999pp} one would anticipate that the $\langle
B\overline{B}\rangle$
correlator is dominated in our $R$-range and for our purposes by
$\langle B\rangle\langle\overline{B}\rangle$.
This is currently being investigated.

\acknowledgements We appreciate useful discussions with S.~G\"usken und
M.~Peardon during early stages of this research.  K.S. thanks
F. Niedermayer for an interesting discussion. This work was supported by
DFG Graduiertenkolleg ``Feldtheoretische und Numerische Methoden in der
Statistischen und Elementarteilchenphysik''.  G.B.\ acknowledges support from
DFG grants Ba 1564/3-1, 1564/3-2 and 1564/3-3 as well as EU grant
HPMF-CT-1999-00353.  The HMC productions were run on an APE100 at INFN Roma.
We are grateful to our colleagues F. Rapuano and G. Martinelli for the fruitful
T$\chi$L-collaboration.  Analysis was performed on the CRAY T3E system of ZAM at
Research Center J\"ulich.

\end{document}